\documentstyle{article}
\newtheorem{rem}{Remark}

\newtheorem{pr}{Proposition}
\begin{document}
\begin{center}
\Large{\bf FOUR-DIMENSIONAL RICCI-FLAT SPACE  \\[2mm] DEFINED BY
THE KP EQUATION
 }\vspace{4mm}\normalsize
\end{center}
 \begin{center}
\Large{\bf Valery Dryuma}\vspace{4mm}\normalsize
\end{center}
\begin{center}
{\bf Institute of Mathematics and Informatics AS Moldova, Kishinev}\vspace{4mm}\normalsize
\end{center}
\begin{center}
{\bf  E-mail: valery@dryuma.com;\quad cainar@mail.md}\vspace{4mm}\normalsize
\end{center}
\begin{center}
{\bf  Abstract}\vspace{4mm}\normalsize
\end{center}

   Four-dimensional affinely connected  Ricci-flat space dependent from solutions of the
   Kadomtsev-Petviashvili equation are constructed.
   Conditions of metrizabilty of corresponding connection are
   investigated. Their properties are discussed.

\section{The Ricci-flat 8-dim metrics}

     An examples of the eight-dimensional Ricci-flat metrics of Riemann extensions of affinely-connected
     four-dimensional spaces $A^4$ dependent from solutions of the KP-equation were obtained recently by
      author (\cite{dryuma7}).

     Here we give the example of  four-dimensional Ricci-flat affinely connected space $E^4$
     defined by solutions of the KP-equation which has Ricci-flat the Riemann extension also
     dependent from solutions of the KP-equation.
\begin{pr}
     Eight-dimensional Riemann space in local coordinates
$(x,y,z,t,P,Q,U,V)$ equipped with the metric
\begin{equation} \label{dryuma:eq1}
{{\it ds}}^{2}=\left (-2\,P{\it H11}(y,z,t)-2\,P{\frac {\partial
}{
\partial t}}F(y,z,t)-2\,{\it H12}(y,z,t)Q-2\it \Gamma^3_{11}(y,z,t)U+2\,{\it H22}(y,z,t)V\right ){{\it
dx}}^ {2}+$$$$+2\,\left (2\,{\it H11}(y,z,t)Q-2\,{\it
H21}(y,z,t)V\right ){\it dx }\,{\it dy}+2\,\left (-2\,\left
({\frac {\partial }{\partial z}}F(y,z, t)\right )V+2\,\left
({\frac {\partial }{\partial t}}F(y,z,t)\right )U \right ){\it
dx}\,{\it dz}+$$$$+2\,{\it dx}\,{\it dP}+\left (2\,{\it H11}(y
,z,t)U-2\,{\it H31}(y,z,t)V\right ){{\it dy}}^{2}+2\,{\it
dy}\,{\it dQ }+2\,{\it dz}\,{\it dU}+2\,{\it dt}\,{\it dV}
\end{equation}
is a Ricci-flat  $$R_{ij}=0$$ if the conditions on the
coefficients $\it Hij(y,z,t)$
\begin{equation} \label{dryuma:eq2}
{\frac {\partial }{\partial y}}{\it H12} \left( y,z,t \right)
-{\frac {\partial }{\partial t}}{\it H22} \left( y,z,t \right) =0
$$ $$ -{\frac {\partial }{\partial y}}{\it H11} \left( y,z,t
\right) + {\frac {\partial }{\partial t}}{\it H21} \left( y,z,t
\right) =0 $$ $$ -{\frac {\partial }{\partial z}}{\it H11} \left(
y,z,t \right) +{\frac {\partial }{\partial t}}{\it H31} \left(
y,z,t \right)=0
\end{equation}
are valid.

  An arbitrary functions $\it \Gamma^3_{11}(y,z,t)$ and $F(y,z,t)$ in so doing satisfy the relation
\[
{\frac {\partial }{\partial z}}\it \Gamma^3_{11}(y,z, t)=2\,\left
({\frac {\partial }{\partial t}}F(y,z,t)\right )^{2}+2\,{ \it
H11}(y,z,t){\frac {\partial }{\partial t}}F(y,z,t)+2\,\left ({\it
H11}(y,z,t)\right )^{2}
\]
or
  \begin{equation} \label{dryuma:eq31}
\it \Gamma^3_{11}(y,z,t)=\int \!2\,\left ({\frac {
\partial }{\partial t}}F(y,z,t)\right )^{2}+2\,{\it H11}(y,z,t){\frac
{\partial }{\partial t}}F(y,z,t)+2\,\left ({\it H11}(y,z,t)\right
)^{2 }{dz}+{\it \_F1}(y,t).
 \end{equation}
\end{pr}

    The  metric (\ref{dryuma:eq1}) is an example of the Riemann extension of
 affinely-connected  four-dimensional space with symmetrical connection
 $\Gamma^i_{jk}(x^l)=\Gamma^i_{kj}(x^l)$ depending from the local
 coordinates
 $x^i$.

 In general case  it is defined by the expression
  \begin{equation} \label{dryuma:eq3}
ds^2=-2\Gamma^i_{jk}\xi_i dx^j dx^k+2 d\xi_k dx^k,
\end{equation}
where $\xi_k$ -are an additional coordinates
(\cite{paterson&walker}).

    After the substitutions of the form
   \begin{equation} \label{dryuma:eq4}
{\it H11} \left( y,z,t \right) =-1/2\,u \left( y,z,t \right),\quad
{\it H12} \left( y,z,t \right) =-1/3\,v \left( y,z,t \right),$$$$
{\it H21} \left( y,z,t \right) =-2/3\,v \left( y,z,t \right) -1/2\,{
\frac {\partial }{\partial t}}u \left( y,z,t \right),$$$$
{\it H31} \left( y,z,t \right) =-3/4\,w \left( y,z,t \right) +3/8\,
 \left( u \left( y,z,t \right)  \right) ^{2}-{\frac {\partial }{
\partial t}}v \left( y,z,t \right) -1/2\,{\frac {\partial ^{2}}{
\partial {t}^{2}}}u \left( y,z,t \right),$$$$
{\it H22} \left( y,z,t \right) =-1/2\,w \left( y,z,t \right) +1/2\,
 \left( u \left( y,z,t \right)  \right) ^{2}-{\frac {\partial }{
\partial t}}v \left( y,z,t \right) -$$$$-1/2\,{\frac {\partial ^{2}}{
\partial {t}^{2}}}u \left( y,z,t \right) +1/2\,{\frac {\partial }{
\partial y}}u \left( y,z,t \right)
\end{equation}
from the conditions  (\ref{dryuma:eq2})
 the famous KP-equation is followed
$$
{\frac {\partial ^{2}}{\partial t\partial z}}u \left( y,z,t \right) -3
/2\, \left( {\frac {\partial }{\partial t}}u \left( y,z,t \right)
 \right) ^{2}-3/2\,u \left( y,z,t \right) {\frac {\partial ^{2}}{
\partial {t}^{2}}}u \left( y,z,t \right) -1/4\,{\frac {\partial ^{4}}{
\partial {t}^{4}}}u \left( y,z,t \right) -3/4\,{\frac {\partial ^{2}}{
\partial {y}^{2}}}u \left( y,z,t \right)=0
$$
or
$$
{\frac {\partial }{\partial t}}\left(\frac{\partial u \left( y,z,t \right)}{\partial z}-3/2\,u \left( y,z,t \right) {\frac {\partial }{
\partial {t}}}u \left( y,z,t \right) -1/4\,{\frac {\partial ^{3}}{
\partial {t}^{3}}}u \left( y,z,t \right)\right)=3/4\,{\frac {\partial ^{2}}{
\partial {y}^{2}}}u \left( y,z,t \right)
$$
\begin{rem}
     A notion of the Riemann extensions of affinely-connected spaces (\cite{paterson&walker})
was used by author for the studying of geometrical problems in
theory nonlinear dynamical systems and in General Relativity
(\cite{dryuma1}-\cite{dryuma7}).

\end{rem}

\begin{rem}
   About presentation of the KP-equation in form of the system (\ref{dryuma:eq2}) see (\cite{konp}).
\end{rem}

\section{Four-dimensional subspace}

      Eight-dimensional metrics (\ref{dryuma:eq3}) is closely related with a properties of
      four-dimensional spaces in local coordinates $x^k$.

     Let us consider an example.

     The full system of geodesic of the metric (\ref{dryuma:eq3}) decomposes into two
     parts.

     The first part  has the form of the linear system of
      equations for the coordinates $(\xi_k=P,Q,U,V)$
\[
\frac{d^2 \xi_k}{ds^2}+A(x^i)\frac{d \xi_k}{ds}+B(x^i)\xi_k=0
\]
     and the second part for local coordinates $x^i=(x,y,z,t)$ is defined by the system of equations
\[
\frac{d^2 x^k}{ds^2}+\it \Gamma^k_{ij}\frac{d
x^i}{ds}\frac{dx^j}{ds}=0.
\]

    In our case it
       takes the form
     \[
{\frac {d^{2}}{d{s}^{2}}}x(s)+\left ({\frac {d}{ds}}x(s)\right
)^{2}{ \it H11}(y,z,t)+\left ({\frac {d}{ds}}x(s)\right
)^{2}{\frac {
\partial }{\partial t}}F(y,z,t)=0,
\]
\[
{\frac {d^{2}}{d{s}^{2}}}y(s)+{\it H12}(y,z,t)\left ({\frac
{d}{ds}}x( s)\right )^{2}-2\,{\it H11}(y,z,t)\left ({\frac
{d}{ds}}x(s)\right ){ \frac {d}{ds}}y(s) =0,
\]
\[
{\frac {d^{2}}{d{s}^{2}}}z(s)+\Gamma^3_{11}(y,z,t )\left ({\frac
{d}{ds}}x(s)\right )^{2}-2\,\left ({\frac {\partial }{
\partial t}}F(y,z,t)\right )\left ({\frac {d}{ds}}x(s)\right ){\frac {
d}{ds}}z(s)-{\it H11}(y,z,t)\left ({\frac {d}{ds}}y(s)\right )^{2}
=0,
\]
\[
{\frac {d^{2}}{d{s}^{2}}}t(s)-{\it H22}(y,z,t)\left ({\frac
{d}{ds}}x( s)\right )^{2}+2\,{\it H21}(y,z,t)\left ({\frac
{d}{ds}}x(s)\right ){ \frac {d}{ds}}y(s)+2\,\left ({\frac
{\partial }{\partial z}}F(y,z,t) \right )\left ({\frac
{d}{ds}}x(s)\right ){\frac {d}{ds}}z(s)+\]\[+{\it H31
}(y,z,t)\left ({\frac {d}{ds}}y(s)\right )^{2} =0.
\]

     From these relations we find the coefficients of  affine connection
     $\it \Gamma^i_{jk}$ of the 4-dimensional subspace in local coordinates
     $(x^i=x,y,z,t)$
\begin{equation}\label{dryuma:eq5}
 \it \Gamma^1_{11}=\it H11+\frac{\partial F}{\partial
t},\quad \Gamma^2_{11}=\it H12,\quad \Gamma^2_{12}=-\it H11, $$$$
 \it \Gamma^3_{11}=\it \Gamma^3_{11},\quad \it
\Gamma^3_{13}=-\frac{\partial F}{\partial t},\quad
\Gamma^3_{22}=-\it H11, $$$$ \it \Gamma^4_{11}=-\it
H22,\quad\Gamma^4_{12}=\it H21,\quad \Gamma^4_{13}=\frac{\partial
F}{\partial z},\quad \Gamma^4_{22}=\it H31.
\end{equation}

   Corresponding the Ricci tensor
\[
R_{ij}=\partial_k \Gamma^k_{ij}-\partial_i \Gamma^k_{kj}+
\Gamma^k_{kl}\Gamma^l_{ij}- \Gamma^k_{im} \Gamma^m_{kj}
\]
    is symmetrical
   \[
   R_{ij}=R_{ji}
   \]
and it is equal to zero $R_{ik}=0$ if the conditions
~(\ref{dryuma:eq2}),~(\ref{dryuma:eq31}) are hold.

    As the example of geodesic equation connected with an additional coordinates we refer
    to the expression on the coordinate $V$
\[
\left (\left ({\frac {d}{ds}}x(s)\right )^{2}{\frac {\partial }{
\partial t}}{\it H11}(y,z,t)+\left ({\frac {d}{ds}}x(s)\right )^{2}{
\frac {\partial ^{2}}{\partial {t}^{2}}}F(y,z,t)\right
)P+\]\[+\left (-2\, \left ({\frac {d}{ds}}x(s)\right )\left
({\frac {d}{ds}}y(s)\right ){ \frac {\partial }{\partial t}}{\it
H11}(y,z,t)+\left ({\frac {d}{ds}}x (s)\right )^{2}{\frac
{\partial }{\partial t}}{\it H12}(y,z,t)\right ) Q+\]\[+{\frac
{d^{2}}{d{s}^{2}}}V(s)-\left ({\frac {d}{ds}}x(s)\right )^{2}
\left ({\frac {\partial }{\partial t}}{\it H22}(y,z,t)\right
)V-\left ({\frac {d}{ds}}y(s)\right )^{2}\left ({\frac {\partial
}{\partial t}} {\it H11}(y,z,t)\right )U+\]\[+2\,\left ({\frac
{d}{ds}}x(s)\right )\left ( {\frac {d}{ds}}y(s)\right )\left
({\frac {\partial }{\partial t}}{\it H21}(y,z,t)\right )V+\left
({\frac {d}{ds}}x(s)\right )^{2}\left ({ \frac {\partial
}{\partial t}}\it \Gamma^3_{11}(y,z,t )\right )U-\]\[-2\,\left
({\frac {d}{ds}}x(s)\right )\left ({\frac {d}{ds}} z(s)\right
)\left ({\frac {\partial ^{2}}{\partial {t}^{2}}}F(y,z,t) \right
)U+\left ({\frac {d}{ds}}y(s)\right )^{2}\left ({\frac {
\partial }{\partial t}}{\it H31}(y,z,t)\right )V+\]\[+2\,\left ({\frac {d}{
ds}}x(s)\right )\left ({\frac {d}{ds}}z(s)\right )\left ({\frac {
\partial ^{2}}{\partial t\partial z}}F(y,z,t)\right )V=0.
\]

    Corresponding equations for the coordinates $U,P,Q$ have more cumbersome
    form and may be omitted here.

\begin{rem}
   As distinct from the example of eight-dimensional the Ricci-flat
   Riemann space related with the KP-equation considered in the  article
   (\cite{dryuma7}) the metric $(\ref{dryuma:eq1})$ has a Ricci-flat affinely connected four-dimensional
   subspace.
\end{rem}

   In this connection it should be noted that problem of metrizability of the
   connection $\it \Gamma^i_{jk}$ defined by the expressions
   $(\ref{dryuma:eq5})$ has of great importance.

   It is known that in order to obtain the metric tensor $g_{ij}$
   correspondent the connection
   $\it \Gamma^i_{jk}$ we have to solve the system of equations
\[
\nabla_i g_{jk}=\partial_i g_{jk}-\it \Gamma^l_{ij}g_{lk}-\it
\Gamma^l_{ik}g_{jl}=0.
\]

   A necessary and sufficient condition of integrability of this
   system implies in the studying of the properties of the set of
   equations
   \[
   \it R^k_{i j l}g_{km}+\it R^k_{i j m}g_{kl}=0,
   \]
\[
\nabla_s \it R^k_{i j l}g_{km}+\nabla_s \it R^k_{i j m}g_{kl}=0,
\]
\[
  .......................................,
\]
where $\it R^k_{i j l}$ are the components of Riemann tensor of
the connection $(\ref{dryuma:eq5})$.

     In the case being considered the number of  components of the Riemann tensor
     is thirty-six and affine connection $(\ref{dryuma:eq5})$ obeys the condition
$ \Gamma^i_{ki}=0.$

    On this basis and taking in consideration the relation
$ \Gamma^i_{ki}=\frac{1}{2g}\frac{\partial g}{\partial x^k} $
there is followed that in the case of metrizability of the
connection $(\ref{dryuma:eq5})$ determinant of corresponding
metric tensor must to be constant.

     \section{ Acknowledgements}

     The research of author was partially supported by Grant CSTD AS
     Moldova and Grant RFBR.


\begin{thebibliography}
\footnotesize

\bibitem{paterson&walker} Paterson E.M. and Walker A.G.,Riemann extensions,
 {\it Quart.J.Math.Oxford},1952, V.3,19--28.

 \bibitem{dryuma1} Dryuma V., The Riemann and Einstein-Weyl geometries in the theory of ODE's
 , their applications and all that,
 {New Trends in Integrability and Partial Solvability, 115-156, (eds.A.B.Shabat et al.) 2004, Kluwer Academic Publishers},
{\it arXiv: nlin: SI/0303023, 11 March, 2003}, 1--37.

\bibitem{dryuma3:dryuma} Dryuma V.,On the Riemann Extension of the G\"odel space -time metric,
{\it Buletinul A.S.R.M, Matematica}, 2005, No.3(49), p.43--62.

\bibitem{dryuma2} Dryuma V.,On solutions of a Heavenly equations and their generalizations,
{\it arXiv:math.DG/0510526 v1, 31 Oct 2006}, p.1--13.

\bibitem{dryuma4:dryuma} Dryuma V.,The Riemann Extension of the Schwarzshild space-time, {\it\it Buletinul A.S.R.M, Matematica},
 2003, No. 3(43), p.92--103.

\bibitem{dryuma5} Dryuma V.,The Riemann Extension of space-time with vanishing curvature invariants,
 {\it http://www.math.kth.se/4ecm/abstract/14.27.pdf}, Sweden, Stockholm, June 27-July 2,~ECM'2004.

\bibitem{dryuma6} Dryuma V., Riemann Extension in theory of the first order systems of differential
equations,
 {\it arXiv:math.DG/0510526 v1, 25 Oct 2005}, p.1--21.

\bibitem{dryuma7} Dryuma V., Eight-dimensional the Ricci-flat
Riemann space related with the KP-equation, {\it arXiv: 0810.0346
v1, 2 Oct 2008}, p.1--5.

\bibitem{konp} B.G.Konopelchenko, Quantum deformations of asociative algebras and integrable systems,
{\it arXiv: 0802.3022 v2 [nlin. SI] 28 Apr, 2008}, p.1--24.

 \end{thebibliography}
 \end{document}